\documentclass[conference]{IEEEtran}
\IEEEoverridecommandlockouts
\usepackage{cite}
\usepackage{amsmath,amssymb,amsfonts}
\usepackage{algorithmic}
\usepackage{graphicx}
\usepackage{textcomp}
\usepackage{hyperref}
\usepackage[ruled,lined,noresetcount ]{algorithm2e}
\usepackage{xcolor}
\def\BibTeX{{\rm B\kern-.05em{\sc i\kern-.025em b}\kern-.08em
    T\kern-.1667em\lower.7ex\hbox{E}\kern-.125emX}}
\usepackage{tikz}
\usetikzlibrary{positioning}
\usepackage{pgfplots}
\pgfplotsset{compat=1.11}

\begin{document}

\title{End-to-End Deep Learning in Phase Noisy Coherent Optical Link
}

\author{\IEEEauthorblockN{Omar Alnaseri\IEEEauthorrefmark{1}, ~\IEEEmembership{Senior Member,~IEEE} and Yassine Himeur\IEEEauthorrefmark{4}, ~\IEEEmembership{Senior Member,~IEEE}
}
\IEEEauthorblockA{\IEEEauthorrefmark{1}Electrical Engineering Department, DHBW Cooperative State University, Ravensburg, Germany, \\
Email: alnaseri.omar@dozent.dhbw-ravensburg.de}
\IEEEauthorblockA{\IEEEauthorrefmark{4}College of Engineering and Information Technology, University of Dubai, Dubai, United Arab Emirates, \\
Email: yhimeur@ud.ac.ae}
}
\maketitle

\begin{abstract}
In coherent optical orthogonal frequency-division multiplexing (CO-OFDM) fiber communications, a novel end-to-end learning framework to mitigate Laser Phase Noise (LPN) impairments is proposed in this paper. Inspired by Autoencoder (AE) principles, the proposed approach trains a model to learn robust symbol sequences capable of combat LPN, even from low-cost distributed feedback (DFB) lasers with linewidths up to 2 MHz. This allows for the use of high-level modulation formats and large-scale Fast Fourier Transform (FFT) processing, maximizing spectral efficiency in CO-OFDM systems. By eliminating the need for complex traditional techniques, this approach offers a potentially more efficient and streamlined solution for CO-OFDM systems. The most significant achievement of this study is the demonstration that the proposed AE-based model can enhance system performance by reducing the bit error rate (BER) to below the threshold of forward error correction (FEC), even under severe phase noise conditions, thus proving its effectiveness and efficiency in practical deployment scenarios.

\end{abstract}

\begin{IEEEkeywords}
Fiber optics communications, CO-OFDM, deep neural network, laser phase noise, autoencoder
\end{IEEEkeywords}

\section{Introduction}
Coherent Optical Orthogonal Frequency-Division Multiplexing (CO-OFDM) is a leading technology for long-haul fiber transmission \cite{roumpos2023high}. This is due to its superior capability to mitigate Polarization-Mode Dispersion (PMD) and Chromatic Dispersion (CD), while maintaining high spectral utilization \cite{b1,srinivasan2023end}. However, the performance of CO-OFDM systems can be severely affected by Laser Phase Noise (LPN), particularly when employing advanced modulation schemes and large-scale Fast Fourier Transform (FFT) processing. To address these challenges, various advancements have been made in Carrier Phase Recovery (CPR) techniques, particularly in the areas of pilot symbol design and estimation algorithms \cite{xing2023end}. 

For instance, \cite{b1a} and \cite{b1b} introduced a novel feedforward carrier recovery algorithm designed for M-ary Quadrature Amplitude Modulation (M-QAM) constellations in an coherent optical receiver. The algorithm stands out because it does not utilize any feedback loop, making it exceptional resilience to LPN. This characteristic is especially beneficial when employing higher-level QAM constellations. Further advancing these methods, \cite{b2} proposed a residual carrier modulation technique that enables efficient recovery of both carrier frequency and phase. This innovative approach improves the bit rate by 41\%, surpassing conventional time-domain pilot techniques, this approach sets a new record for the product of laser linewidth and symbol duration. The scheme is particularly advantageous for MHz linewidth Distributed Feedback (DFB) lasers, which are commonly used in affordable coherent optical communications.

Building on this foundation, \cite{b3} explored various blind phase noise detection methods for CO-OFDM transmission, introducing a decision-direct-free method with a testing process consisting of three phases. This method demonstrates similar performance to conventional techniques but with reduced complexity. Additionally, the researchers proposed a novel cost function aimed at enhancing the efficiency of phase noise compensation. 

Despite these innovations, RF pilot tones remain a popular approach for mitigating LPN. In \cite{b5}, the performance of a system employed an RF-pilot-based method for mitigation LPN was compared to the traditional common-phase error method. The results showed that the RF-pilot-based scheme significantly enhances the tolerable laser linewidth. For example, in a 112-Gb/s transmission, the laser linewidth tolerance was ten times higher than that of traditional common-phase error compensation methods. Furthermore, \cite{b6} and \cite{b7} proposed a Sub-carrier-Index Modulation OFDM (SIM-OFDM) employing the RF-pilot, which combats both LPN and fiber nonlinearities in CO-OFDM transmission with a large-scale FFT of 1024 and 16-QAM modulation form. SIM-OFDM revealed a higher tolerance to LPN than conventional CO-OFDM and achieves better performance in terms of Bit Error Rate (BER) and optical Signal-to-Noise Ratio (OSNR) compared to techniques like self-cancellation and Partial Carrier Filling (PCF). Moreover, \cite{b8} investigated the correlation between phase noise in the symbol domain of Discrete Fourier Transform (DFT)-Spread CO-OFDM transmission and its effects in the time domain. The study proposed a DFT-Spread OFDM-based phase noise compensation method, specifically designed for higher-level QAM transmission with large-scale FFT processing. The simulation of a dual polarization CO-OFDM system with 64-QAM over 80 km illustrated the effective of using PNC in mitigating Inter-Carrier Interference (ICI) induced by LPN.
The findings suggest that the PNC scheme improves laser linewidth tolerance in CO-OFDM transmission, 
Providing a promising solution for short-distance, high bit rate, and high-level QAM coherent transmission. 

\begin{figure*}[htbp]
    \centering
    \includegraphics[width=1.0\linewidth]{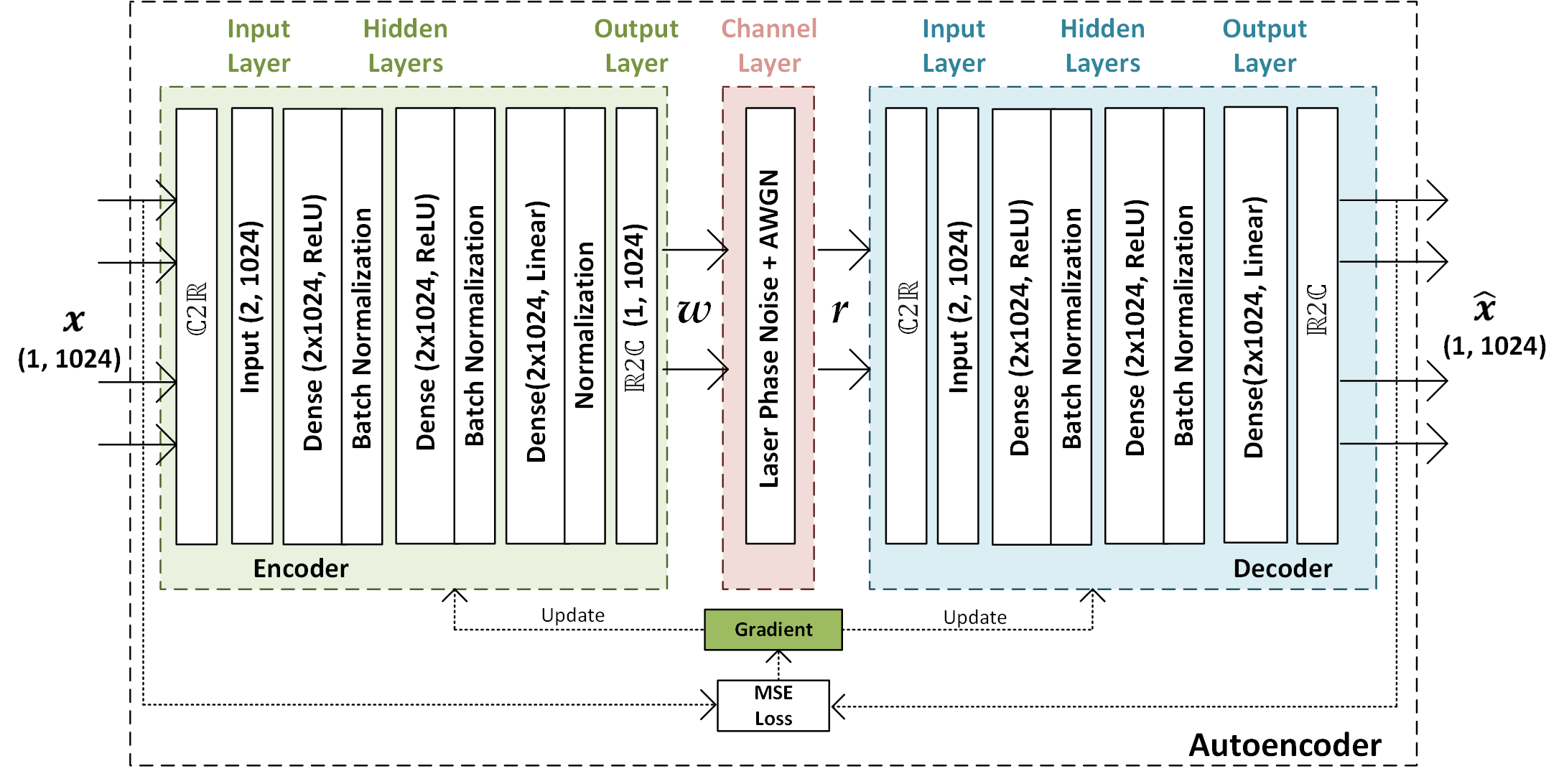}
    \caption{AE end-to-end learning structure}
    \label{fig1}
\end{figure*}

While traditional methods continue to dominate communication systems, they typically rely on a chain of separate transmitter and receiver signal processing blocks \cite{jian2022reconfigurable}. This segmented approach often leads to sub-optimal module performance, ultimately limiting the achievable information rates \cite{da2024survey}. In contrast, Deep Learning (DL)-based communication systems, which are inspired by Autoencoder (AE) architectures, have garnered significant attention in recent years \cite{b9,b10}. These systems offer the potential to optimize end-to-end communication performance by leveraging the ability of deep learning models to learn robust latent representations. Specifically, AEs can effectively capture the characteristics of the communication channel, as well as the impacts of noise and interference. Recently, \cite{marasinghe2024constellation} investigated the challenge of phase noise (PN) impairment in sub-THz communications. Employing a Wiener phase noise model and stringent PAPR constraints, would lead to some  limitations in constellation design flexibility, potentially leading to suboptimal performance.

Building upon this concept, \cite{b11} proposed an end-to-end learning for OFDM systems operating in multipath channels. Their study demonstrated promising results for QPSK modulation and small FFT sizes. However, the exploration of DL-based solutions for achieving higher spectral efficiency—characterized by larger FFT sizes and advanced modulation formats—remains relatively unexplored. Recent research by \cite{b12} has highlighted the effectiveness of AEs in mitigating LPN generated by low-cost DFB lasers (with linewidths exceeding 1 MHz) in a transmission system utilizing 16-QAM and a 1024-FFT size. Remarkably, these models have outperformed traditional RF pilot-based techniques \cite{b13, b14}. Nevertheless, it is worth noting that the model presented in \cite{b12} only accounted for a channel layer of Additive White Gaussian Noise (AWGN) during training, limiting its general applicability.

In this paper, we aim to address the pressing challenge of mitigating high LPN in CO-OFDM fiber communications, particularly when employing high-level modulation formats and large-scale FFT processing. To achieve this, we propose a novel AE-based communication model that is specifically tailored for Inverse Fast Fourier Transform (IFFT)/FFT of1024-point, and 16-QAM format. By training the AE on a random walk phase noise channel to demonstrate exceptional resilience to LPN, we effectively address the regression problem inherent in this application.
Notably, this approach eliminates the need for additional LPN mitigation techniques, offering a significant advantage over traditional RF-pilot-based methods by reducing complexity and potential performance bottlenecks. Our proposed method presents a promising solution for achieving robust CO-OFDM transmission in practical fiber optic environments. The key contributions of this paper are as follows:
\begin{itemize}
    \item This model is specifically designed for 1024-point FFT-size, 16-QAM CO-OFDM fiber transmission, addressing the challenge of LPN in high-level modulation formats and large-scale FFT systems.
    
    \item By training the model on a random walk phase noise channel, the proposed solution demonstrates exceptional tolerance to LPN, eliminating the need for additional mitigation techniques such as RF-pilot-based methods.
    
    \item The model offers a significant advantage over traditional techniques by reducing the complexity and potential performance bottlenecks, making it a practical solution for robust CO-OFDM transmission in real-world fiber optic environments.
\end{itemize}

\section{OFDM-Autoencoder Based Communication Systems}
AEs are a class of deep learning neural networks primarily designed for data reconstruction, offering a novel and powerful approach to enhancing communication systems \cite{b9}. In the context of communication, the AE framework enables the transmitter to function as an encoder, generating robust latent representations that are resistant to various channel impairments. The receiver, acting as a decoder, is responsible for accurately reconstructing the original data from these latent representations. This end-to-end training paradigm optimizes together the transceiver and the receiver, ensuring efficient data transmission across noisy or impaired channels. A key advantage of this approach is the encoder's ability to learn and generate resilient latent representations, which are encoded in its hidden layers. Once trained, the transmitter and receiver can operate independently while leveraging the learned weights and layer structures to ensure robust performance.

As depicted in Figure~\ref{fig1}, To process complex-valued data in neural networks, a concatenation operation is employed to separate the real and imaginary components into two distinct real-valued representations ($\mathbb{R}2\mathbb{C}$). The output from the network is then converted back to complex form using the inverse operation($\mathbb{C}2\mathbb{R}$). The proposed AE architecture comprises two dense layers employing Rectified Linear Unit (ReLU) activation functions. Each dense layer in both the encoder and decoder is followed by a batch normalization layer. This batch normalization speeds up AE training and reduces overfitting \cite{ioffe2015batch}. The final layer employs a linear activation function, which generates numerical data representations. The encoder is responsible for producing a latent representation, denoted as $w$, which is passed through a random walk phase noise channel. The output from this noisy channel, represented by $r$, is then fed into the decoder, where the original data $x$ is reconstructed. This approach enables efficient mitigation of channel impairments, such as phase noise, thereby enhancing the overall performance of the communication system.

\begin{table*}[]
\caption{AE Training Hyperparameters}
\label{tab1}
\centering
\begin{tabular}{|l|ll|}
\hline
\textbf{Hyperparameter} & \multicolumn{1}{l|}{\textbf{Description}}                                  & \textbf{Value} \\ \hline
Batch size              & \multicolumn{1}{l|}{The batch size used for training}                  & N              \\ \hline
Optimizer               & \multicolumn{1}{l|}{Optimizer algorithm}                  & Adam              \\ \hline
ReduceLROnPlateau       & \multicolumn{2}{l|}{A callback mechanism to reduce the learning rate when a metric plateaus.}                                                           \\ \hline
Learning rate           & \multicolumn{1}{l|}{The minimum learning rate}                             & $1\cdot 10^{-10}$          \\ \hline
Factor                  & \multicolumn{1}{l|}{The learning rate reduction factor} & 0.1            \\ \hline
Patience                & \multicolumn{1}{l|}{\begin{tabular}[c]{@{}l@{}}Patience parameter\end{tabular}} & 10 \\ \hline
EarlyStopping           & \multicolumn{2}{l|}{Early stopping mechanism}          \\ \hline
Min delta               & \multicolumn{1}{l|}{Minimum change to consider as improvement}             & 0.0001         \\ \hline
Patience                & \multicolumn{1}{l|}{\begin{tabular}[c]{@{}l@{}}Patience parameter\end{tabular}}  & 50 \\ \hline
\end{tabular}
\end{table*}

\begin{table*}[]
\label{tab2}
\centering
\caption{Simulation Parameters}
\begin{tabular}{|l|l|l|}
\hline
\textbf{Parameter} & \textbf{Description} & \textbf{Values} \\ \hline
FFT Size & Size of the iFFT/FFT used in the system & 1024 \\ \hline
Modulation Format & Type of modulation used & 16-QAM \\ \hline
Laser Linewidth & Range of laser linewidths considered & 10 kHz, 100 kHz, 200 kHz, 500 kHz, 1 MHz, 2 MHz, 3 MHz \\ \hline
Baud Rate & Symbol rate of the system & 32 GHz \\ \hline
Noise Model & Type of noise model used for simulation & Random Walk Phase Noise \\ \hline
AE models & Trained AE model under linewidths & 10 kHz, 100 kHz \\ \hline
\end{tabular}
\end{table*}

\subsection{Laser Phase Noise Layer}
To simulate LPN, random noise samples with a specified variance are added to the oscillator's phase at each time step \cite{b15}. This accumulation of noise over time models the random-walk phase noise process, which is a significant impairment in coherent optical communication systems. The discrete-time random-walk model is generated iteratively for $N$ samples, can be defined as:

\begin{equation}
\label{eq:1}
\theta_{i+1} = \theta_{i} + C, \quad i \in \{0, N-1\},
\end{equation}
where $\theta_{i}$ is the phase noise at sample $i$, and $\theta_{i+1}$ represents the phase noise at the next time step. The random variable $C$ follows a normal distribution with mean $\mu$ and standard deviation $\sigma$, defined as:

\begin{equation}
\label{eq:2}
C \sim \mathcal{N}(\mu, \sigma^2).
\end{equation}

This random variable follows a zero-mean normal distribution, where the standard deviation $\sigma$ is directly related to the laser linewidth $\Delta v$ and the sampling period $T_{s}$:

\begin{equation}
\label{eq:3}
\sigma^{2} = 2 \pi \cdot \Delta v \cdot T_{s}.
\end{equation}

The diffusion coefficient of the Wiener process is also linked to the variance, where $\sigma^{2} = D_{coeff} \cdot t$. This diffusion coefficient, $D_{coeff}$, dictates how rapidly the random walk spreads over time. Thus, it is an important key in characterizing the behavior of the random-walk phase noise. A higher laser linewidth increases the diffusion coefficient, leading to faster phase noise spread over time. Similarly, a reducing sampling period leads to a larger diffusion coefficient, further emphasizing the impact of these parameters on phase noise dynamics.

\subsection{Autoencoder Training}
The goal of this work is to train an AE to mitigate the effects of LPN in CO-OFDM systems, specifically by reconstructing 16-QAM in the presence of random walk phase noise. The training process focuses on discovering resilient symbol representations $w$ that can effectively counteract the phase noise impact. As illustrated in Figure~\ref{fig1}, the proposed end-to-end learning architecture begins with an input layer of dimension $N$, equivalent to the iFFT size. This is followed by dense encoder layers that utilize ReLU activation and a batch normalization, extracting features of dimension $2N$. The phase noise channel, characterized by different linewidths $\Delta v$ (e.g., 10 kHz and 100 kHz), is introduced at the channel layer, which remains non-trainable. The baud rate ($T_{s}$) is set to 32 GHz, and the decoder mirrors the encoder structure, concluding with a linear output layer of dimension $N$.

In this architecture, the AE is trained to reconstruct 16-QAM symbols despite the occurrence of a random walk phase noise channel. The model's objective is to learn robust symbol representations $w$ that are resilient to phase noise impairments. As training progresses, the encoder and decoder layers leverage ReLU activation and a batch normalization to enhance the extraction and reconstruction processes. The non-trainable phase noise channel layer introduces varying linewidths (10 kHz and 100 kHz), simulating real-world conditions where LPN can significantly impair system performance. The detailed training hyperparameters are presented in Table~\ref{tab1}.

Figure~\ref{fig2} shows the convergence behavior of the AE during training for various levels of LPN. The mean squared error (MSE) is considered as a loss function, that is defined by:

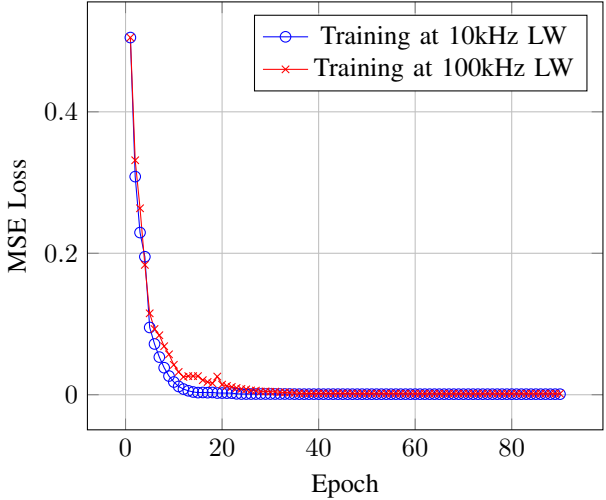
\begin{figure}[htbp]
\begin{tikzpicture}
    \begin{axis}[grid=major,xlabel=Epoch, ylabel=MSE Loss, legend pos=north east]
        \addplot[blue, mark=o] table[x=e, y=10k] {loss.dat};
        \addlegendentry{Training at 10kHz LW}
        \addplot[red, mark=x] table[x=e, y=100k] {loss.dat};
        \addlegendentry{Training at 100kHz LW}
    \end{axis}
\end{tikzpicture}
\caption{Loss function convergence (MSE)}
\label{fig2}
\end{figure}

\begin{equation}
\label{eq:M1}
\mathcal{L}_{MSE} = \frac{1}{N} \sum^{N}_{i=1} |\hat{x} - x|^{2},
\end{equation}
where $N$ represents the number of samples, and $x$ and $\hat{x}$ are the original and estimated 16-QAM symbols, respectively. The results reveal that the model converges to a loss of approximately 0.001 within 24 epochs for a 10 kHz linewidth and 37 epochs for a 100 kHz linewidth. This observation suggests that higher phase noise levels require slightly more training time to achieve convergence, highlighting the model's adaptability to different noise environments.

The training results confirm the effectiveness of proposed AE in mitigating LPN, by successfully learning to reconstruct transmitted symbols even in challenging noise conditions. This achievement underscores the potential of deep learning-based solutions for addressing complex impairments in communication systems.

\subsection{Algorithm of the proposed method}
Algorithm \ref{algo1} outlines an end-to-end AE-based approach to combat LPN in CO-OFDM communication systems. The process begins with mapping the input data into a latent representation using dense layers employing ReLU activation and a batch normalization. This encoded data is then passed through a simulated channel where random walk phase noise is applied, representing the LPN experienced in real-world transmission. The noisy data is fed into a decoder, which reconstructs the original data using a similar dense-layer structure. The performance of the model is optimized by minimizing the MSE between the original input and the estimated output, using the Adam optimizer. By training on this setup, the AE learns to mitigate the phase noise effectively, offering an efficient alternative to traditional CO-OFDM systems.

\begin{algorithm}[t!]
\SetAlgoLined 

\KwIn{Transmitted data $x$, FFT size $N$, Laser linewidth $\Delta v$}
\KwOut{Reconstructed data $\hat{x}$}

\textbf{Step 1: Encoder}

Encode perform the mapping of input data $x$ into latent representation $w$ using dense layers (ReLU activation).

Normalize the encoded data.

\textbf{Step 2: Channel with Phase Noise}

Apply random walk phase noise to the latent representation $w$. 

Simulate the phase noise according to:

\[
\theta_{i+1} = \theta_{i} + C, \quad C \sim \mathcal{N}(0, \sigma^2),
\]

where $\sigma^2 = 2\pi \cdot \Delta v \cdot T_s$.

\textbf{Step 3: Decoder}

Feed the phase-noisy representation $r$ into the decoder.

Reconstruct the original data $\hat{x}$ using dense layers with ReLU activation followed by a linear output layer.

\textbf{Step 4: Loss Computation and Optimization}

Compute the mean squared error (MSE) between the original and reconstructed data:

\[
\mathcal{L}_{MSE} = \frac{1}{N} \sum_{i=1}^{N} |x - \hat{x}|^2.
\]

Optimize the AE model using the Adam optimizer to minimize the MSE loss.
\caption{AE-Based CO-OFDM Communication System}
\label{algo1}

\end{algorithm}

\section{Simulation Setup}
\begin{figure}[htbp]
\begin{tikzpicture}

\node[draw,text width=1.3cm,minimum height=1cm,minimum width=1cm] (AE) at 
     (0,0) {AE Encoder};
\node[draw,text width=0.7cm,minimum height=1cm,minimum width=1cm] (iFFT) at 
     (1.7,0) {iFFT};
     
\node[draw,circle,minimum size=0.6cm,fill=white!50] (sum) at (3,0){};
\draw (sum.north east) -- (sum.south west)
    (sum.north west) -- (sum.south east);
\draw (sum.north east) -- (sum.south west)
(sum.north west) -- (sum.south east);

\node[draw,circle,minimum size=0.6cm,fill=white!50] (sum2) at (3.8,0){};
\draw (sum2.north) -- (sum2.south)
    (sum2.east) -- (sum2.west);

\node[draw,text width=0.7cm,minimum height=1cm,minimum width=1cm] (AE_dec) at 
     (5.5,0) {FFT};
\node[draw,text width=1.3cm,minimum height=1cm,minimum width=1cm] (FFT) at 
     (7.2,0) {AE decoder};
\draw[-stealth] (AE.east) |- (iFFT.west) 
    node[near end,above]{};
    
\draw[-stealth] (iFFT.east) |- (sum.west) 
    node[near end,above]{$w$};
\draw[-stealth] (sum.east) |- (sum2.west);
\draw[-stealth] (sum2.east) |- (AE_dec.west) 
    node[near end,above]{$r$};
\draw[-stealth] (AE_dec.east) |- (FFT.west);

\node (v0) [above of=sum, yshift=0.25in] {$e^{j\theta}$};
\draw[-stealth] (v0) -- (sum);

\node (v1) [above of=sum2, yshift=0.25in] {$n$};
\draw[-stealth] (v1) -- (sum2);

\end{tikzpicture}
\caption{CO-OFDM Simulation Setup}
\label{fig3}
\end{figure}
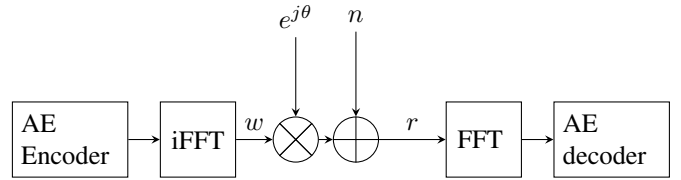

To integrate AE parts into a CO-OFDM transmission system, the AE encoder is positioned at the transmitter as shown in Figure~\ref{fig3}. An IFFT of 1024-point is then applied to the output of encoder(latent space), with each output mapped to a distinct OFDM subcarrier. The transmitted signal is subject to channel impairments, including various linewidths and Additive White Gaussian Noise (AWGN). At the receiver, an FFT of 1024-point FFT is applied to the received signal, transforming it into the time domain. The decoder subsequently reconstructs the original data transmitted by the encoder.

\vspace{0.5cm}

\section{Simulation Results}
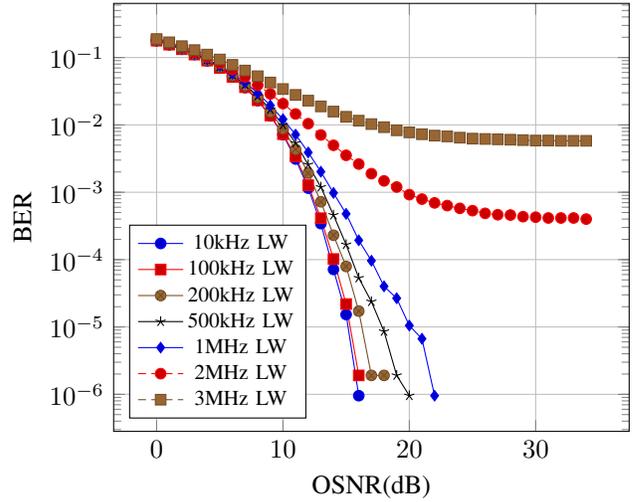
\begin{figure}[htbp]
\begin{tikzpicture}
    \begin{semilogyaxis}[grid=major,xlabel=OSNR(dB), ylabel=BER, legend style={nodes={scale=0.8, transform shape}}, legend pos=south west]
        \addplot table[x=OSNR, y=10k] {BER_10k.dat};
        \addlegendentry{10kHz LW}
        \addplot table[x=OSNR, y=100k] {BER_10k.dat};
        \addlegendentry{100kHz LW}        
        \addplot table[x=OSNR, y=200k] {BER_10k.dat};
        \addlegendentry{200kHz LW}
        \addplot table[x=OSNR, y=500k] {BER_10k.dat};
        \addlegendentry{500kHz LW}
        \addplot table[x=OSNR, y=1M] {BER_10k.dat};
        \addlegendentry{1MHz LW}
        \addplot table[x=OSNR, y=2M] {BER_10k.dat};
        \addlegendentry{2MHz LW}
        \addplot table[x=OSNR, y=3M] {BER_10k.dat};
        \addlegendentry{3MHz LW}
    \end{semilogyaxis}
\end{tikzpicture}
\caption{OSNR vs. BER of AE trained at 10 kHz laser linewidth}
\label{fig4}
\end{figure}

\begin{figure}[htbp]
\begin{tikzpicture}
    \begin{semilogyaxis}[grid=major,xlabel=OSNR(dB), ylabel=BER, legend style={nodes={scale=0.8, transform shape}}, legend pos=south west]
        \addplot table[x=OSNR, y=10k] {BER_100k.dat};
        \addlegendentry{10kHz LW}
        \addplot table[x=OSNR, y=100k] {BER_100k.dat};
        \addlegendentry{100kHz LW}        
        \addplot table[x=OSNR, y=200k] {BER_100k.dat};
        \addlegendentry{200kHz LW}
        \addplot table[x=OSNR, y=500k] {BER_100k.dat};
        \addlegendentry{500kHz LW}
        \addplot table[x=OSNR, y=1M] {BER_100k.dat};
        \addlegendentry{1MHz LW}
        \addplot table[x=OSNR, y=2M] {BER_100k.dat};
        \addlegendentry{2MHz LW}
        \addplot table[x=OSNR, y=3M] {BER_100k.dat};
        \addlegendentry{3MHz LW}
    \end{semilogyaxis}
\end{tikzpicture}
\caption{OSNR vs. BER of AE trained at 100 kHz laser linewidth}
\label{fig5}
\end{figure}
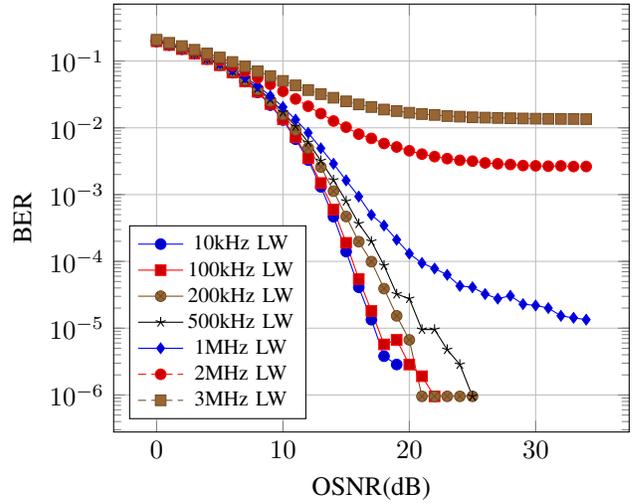

Table~\ref{tab2} shows the simulation parameters used in this paper. To assess the resilience of the proposed AE to LPN, simulations were conducted using various linewidths ranging from 10 kHz to 3 MHz. Both AEs, trained under 10 kHz (Figure~\ref{fig4}) and 100 kHz (Figure~\ref{fig5}) LPN conditions, demonstrated effective mitigation up to 2 MHz. However, neither model could adequately handle 3 MHz linewidths. Interestingly, the AE trained under a 10 kHz linewidth exhibited superior performance at 1 MHz over the AE trained under a 100 kHz linewidth, as BER of a system employed a 1 MHz linewidth experiences an error floor in Figure~\ref{fig5} while not in Figure~\ref{fig4}.

\begin{figure}[htbp]
\begin{tikzpicture}
    \begin{axis}[grid=major,xlabel=Linewidth(Hz), ylabel=Required OSNR(dB), legend pos=north west]
        \addplot[blue, mark=o] table[x=lw, y=10k] {lw.dat};
        \addlegendentry{Training at 10kHz LW}
        \addplot[red, mark=x] table[x=lw, y=100k] {lw.dat};
        \addlegendentry{Training at 100kHz LW}
    \end{axis}
\end{tikzpicture}
\caption{OSNR required to achieve FEC threshold vs. laser linewidth}
\label{fig6}
\end{figure}
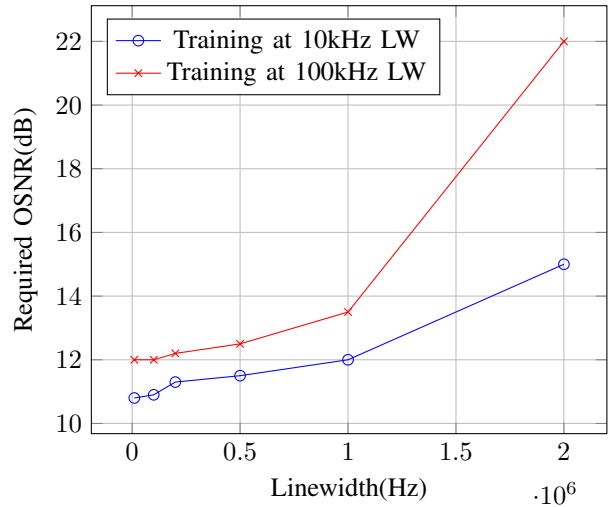

Figure~\ref{fig6} illustrates the OSNR required to achieve the FEC threshold for various laser linewidths. Both AEs, trained under 10 kHz and 100 kHz conditions, demonstrate resilience to linewidth up to 2 MHz. Notably, the AE trained with a 10 kHz linewidth consistently outperforms the 100 kHz model. This suggests that for CO-OFDM systems employing 16-QAM modulation and IFFT/FFTs of 1024-point, the 10 kHz-trained AE offers a more robust solution for mitigating linewidth up to 1 MHz.

\section{Conclusions}
This research introduces a novel AE-based approach for CO-OFDM fiber transmission that offers exceptional resilience to LPN without relying on traditional mitigation techniques. By eliminating the complexity and potential performance bottlenecks associated with conventional methods, this proposed approach paves the way for simplified communication systems. This approach demonstrates remarkable tolerance to Laser Phase Noise (LPN), effectively handling linewidths up to 2 MHz which makes it ideal for compensating for DFB lasers in high-level modulation formats and large FFT sizes, where spectral efficiency is paramount. The results demonstrate that using AE components trained with a 10 kHz laser linewidth achieve exceptional BER performance, offering a viable alternative to computationally demanding conventional phase noise mitigation techniques. Training the proposed AE model with a channel layer simulating a 10 kHz laser linewidth remarkably enhances its robustness to LPN variations, particularly compared to that AE model simulating a 100 kHz linewidth. This finding underscores the criticality of carefully designing the AE model to achieve effective performance across diverse laser linewidth conditions. Future work should investigate two points: (1) combining AEs with techniques like RF pilot tones or digital signal processing algorithms might provide enhanced performance, (2) investigating the application of AE-based techniques to other types of impairments in CO-OFDM systems.


\bibliographystyle{IEEEtran}
\bibliography{references}

\end{document}